\documentclass{article}

\usepackage{arxiv}
\usepackage[pdftex]{graphicx}
\graphicspath{{./figures/}}
\usepackage{booktabs}
\usepackage{siunitx}
\usepackage{amsmath,amssymb,amsfonts}
\usepackage{algorithmic}
\usepackage{textcomp}
\usepackage{xcolor}
\usepackage{subcaption}
\usepackage{cite}
\usepackage{wrapfig}
\usepackage{float}
\usepackage{stfloats}
\usepackage{booktabs}
\usepackage{subcaption}
\allowdisplaybreaks

\DeclareMathOperator*{\argmin}{arg~min}

\DeclareMathOperator{\tr}{tr}

\usepackage{colortbl}
\usepackage{makecell}

\title{On the Unification of Optimal Current Reference Theory for Wound Rotor Synchronous Machines}

\author{
Maxfield Parson-Scherban \\
Department of Electrical Engineering \\
Columbia University \\
New York, NY 10027 \\
\texttt{m.parson@columbia.edu}
\And
Kasra Fallah \\
Department of Electrical Engineering \\
Columbia University \\
New York, NY 10027 \\
\texttt{kasra.fallah@columbia.edu}
\And
Navid Rahbariasr \\
Department of Electrical Engineering \\
Columbia University \\
New York, NY 10027 \\
\texttt{nr3051@columbia.edu}
\And
Bernard Steyaert \\
Department of Electrical Engineering \\
Columbia University \\
New York, NY 10027 \\
\texttt{bernard.steyaert@columbia.edu}
\And
James Anderson \\
Department of Electrical Engineering \\
Columbia University \\
New York, NY 10027 \\
\texttt{james.anderson@columbia.edu}
\And
Matthias Preindl \\
Department of Electrical Engineering \\
Columbia University \\
New York, NY 10027 \\
\texttt{matthias.preindl@columbia.edu}
}
\begin{document}
\maketitle
\begin{abstract}
Controllers for motor drives typically require a current reference which will satisfy the requested torque subject
to system constraints. This work generalizes existing current reference theory to the case of the Wound Rotor Synchronous Machine (WRSM). By incorporating the additional rotor-current degree-of-freedom, along with magnetic saturation, cross-coupling, and speed-dependent core losses, the problem of finding an optimal current reference is formulated within affine flux regions as a quadratically constrained quadratic program using a piecewise-affine approximation derived from finite-element data. The solution is characterized according to the active constraint regime, yielding closed-form or low-dimensional polynomial solutions in several cases, and a small semidefinite program in the voltage constrained regime. The proposed framework extends unified optimal current reference theory beyond the permanent-magnet setting to three degree-of-freedom WRSMs while remaining computationally tractable. Results on a physical WRSM prototype illustrate the effectiveness of the approach across the torque-speed operating envelope.
\end{abstract}


\section{INTRODUCTION}
Electric motors account for a substantial share of global electricity consumption~\cite{waideEnergyEfficiencyPolicyOpportunities2011, GlobalEVOutlook2025}, making efficiency-oriented motor control an important practical problem. In electric vehicle applications, this problem is particularly challenging because traction motors must operate efficiently over a wide range of torque and speed conditions, rather than near a single nominal operating point. This motivates the study of real-time current control strategies that explicitly account for efficiency and feasibility across the operating envelope.

\begin{figure}[!t]
    \centering
    \includegraphics[width=.6\columnwidth]{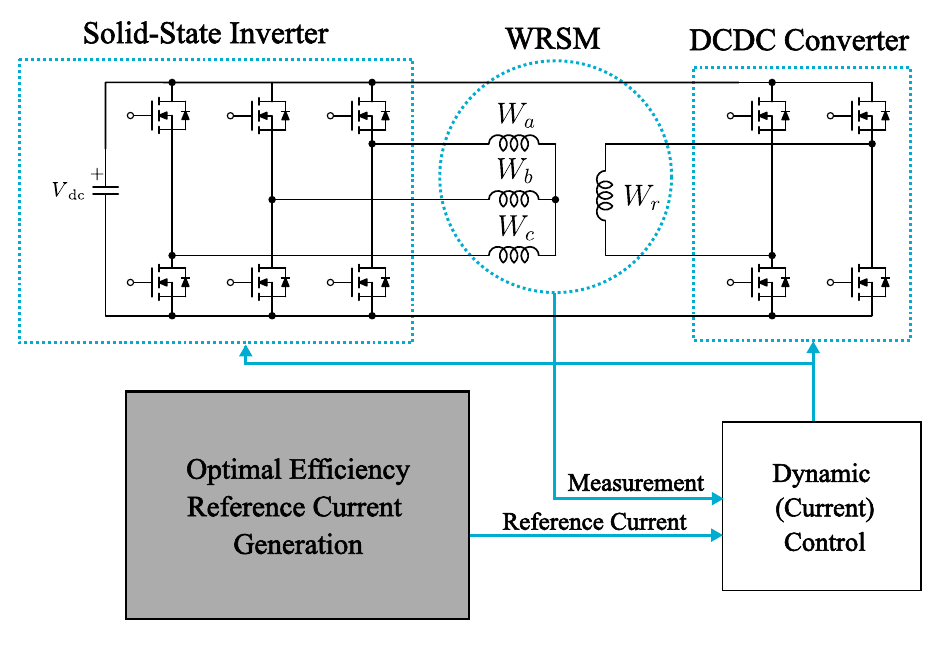}
    \caption{Example synchronous machine (WRSM) control block diagram and schematic.}
    \label{fig:blk}
\end{figure}

A central component in this setting is the optimal current reference generator: for a given torque request and operating speed, it selects stator and rotor current references that satisfy the requested torque, respect current and voltage limits, and minimize electrical losses. This is inherently a constrained decision problem. For a fixed torque-speed pair, the set of feasible current vectors is generally non-unique and, in general, infinite~\cite{morimotoExpansionOperatingLimits1990a}. As illustrated in Fig.~\ref{fig:blk}, current reference generation therefore acts as an optimization layer within the control loop: the reference generator selects the steady-state target, while the lower-level current controller drives the machine to that operating point. The steady-state efficiency of the closed-loop system consequently depends on the quality of this upstream decision.
Optimization-based formulations of motor control problems have received growing attention in both the motor drives and control communities~\cite{batlleSimultaneousIDApassivitybasedControl2008, moehleOptimalCurrentWaveforms2015a, moehleMaximumTorquepercurrentControl2016, doria-cerezoSlidingModeControl2009}. For the Permanent Magnet Synchronous Machine (PMSM), the optimal current reference problem is by now well understood: it admits a Quadratically Constrained Quadratic Programming (QCQP) formulation, and analytical characterizations are available for the relevant operating regimes, including field-weakening~\cite{morimotoExpansionOperatingLimits1990a, preindlOptimalStateReference2015a, eldeebUnifiedTheoryOptimal2018, englertNonlinearModelPredictive2020}.

The Wound Rotor Synchronous Machine (WRSM) has recently re-emerged as a promising alternative for traction applications~\cite{desantiagoElectricalMotorDrivelines2012b}. By replacing permanent magnets with an externally excited rotor winding, the WRSM reduces dependence on rare-earth materials, improves safety under certain fault scenarios, and enables active flux control over a broad operating range~\cite{rossiWoundRotorSalient2006b}. This additional flexibility, however, significantly complicates the current reference generation problem. In contrast to the PMSM case, the rotor current introduces a third control degree-of-freedom, which enlarges the feasible set and makes the structure of the optimal solution substantially more involved, especially in the presence of magnetic saturation, cross-saturation, and speed-dependent core losses.

Existing approaches to WRSM current reference generation are incomplete in complementary ways. Numerical optimization methods can provide high-quality solutions offline~\cite{mullerApplyingMeasurementBasedIron2020}, but their computational cost and runtime variability can limit direct use in embedded real-time implementations. Learning-based surrogates trained on such solutions can improve runtime performance~\cite{monzenArtificialNeuralNetwork}, but they generally do not preserve analytical structure or provide certifiable optimality properties, and may require retraining when machine characteristics change. On the analytical side, the WRSM current reference problem has been formulated as a QCQP in~\cite{parson-scherban2026sdp, parson-scherbanEffectsMagneticSaturation2024a}, and KKT-based analysis of selected operating regimes appear in~\cite{bredaExtendedMTPAFluxWeakening2023a, reinhardOptimalCurrentSetpoint2022}. However, existing analytical treatments either neglect saturation and cross-saturation, or impose assumptions that exclude practically relevant magnetic coupling effects. To the best of our knowledge, a general analytical treatment of the WRSM reference generation problem that simultaneously accounts for saturation, cross-saturation, and core losses across the relevant operating regimes is \textit{not} yet available.

This paper addresses that gap. Using a Piecewise-Affine (PWA) approximation of the nonlinear flux map obtained from finite-element data, we formulate the WRSM current reference generation problem as a QCQP on each affine region. We then characterize the solution structure according to the binding constraints at optimality. This regime-wise decomposition yields closed-form or low-dimensional polynomial solution procedures in several practically relevant cases, and a small semidefinite programming (SDP) relaxation in the voltage-constrained regime. Across the tested operating envelope, the resulting method matches the solution quality of a state-of-the-art nonlinear solver while achieving substantially lower runtime. A comparison with related work is reported in Table~\ref{tab:related_work}.

\newpage
\medskip
\noindent\textbf{Contributions.} The main contributions of this work are as follows.

\smallskip
\noindent $\bullet$ \textbf{WRSM reference-generation formulation.}
We formulate the optimal current reference generation problem for the WRSM as a QCQP on each region of a PWA magnetic model, incorporating saturation, cross-saturation, and a quadratic core-loss model with speed-dependent iron losses.

\smallskip
\noindent $\bullet$ \textbf{Regime-wise analytical solution structure.}
We characterize the active-constraint regimes of the problem and derive closed-form or low-dimensional polynomial solution procedures for four of the resulting cases not handled in \cite{eldeebUnifiedTheoryOptimal2018}. The remaining voltage-constrained case is handled through a small SDP relaxation, which is observed to be exact throughout the tested operating envelope.

\smallskip
\noindent $\bullet$ \textbf{Computationally efficient online reference generation.}
The proposed regime-aware solution strategy exploits problem structure to avoid relying exclusively on generic nonlinear optimization at runtime, making it well suited for real-time reference generation.

\smallskip
\noindent $\bullet$ \textbf{Validation on a prototype WRSM.}
Using finite-element data from a prototype WRSM, we show that the proposed framework reproduces the optimal reference solutions obtained by a state-of-the-art nonlinear solver while reducing median solve time by approximately 95\%, with median runtime below 1\,ms.

\begin{table}[h]
\vspace{8pt}
\centering
\caption{Comparison of Related Work}
\label{tab:related_work}
\renewcommand{\arraystretch}{1.2}
\renewcommand\theadfont{\bfseries}
\renewcommand\theadgape{}
\definecolor{proposedgray}{gray}{0.88}
\setlength{\tabcolsep}{4pt}
\begin{tabular}{lcccccc}
\toprule
\thead{Work} & \thead{Machine} & \thead{Closed\\form} & \thead{Flux\\offset} & \thead{Cross-\\coupling} & \thead{Core\\loss} & \thead{SDP$^\dagger$} \\
\midrule
\cite{preindlOptimalStateReference2015a}  & PMSM & \checkmark & \checkmark & $\times$   & $\times$   & ---        \\
\makecell[l]{\cite{eldeebUnifiedTheoryOptimal2018}\,+\,\cite{hacklGenericLossMinimization2021}} & PMSM & \checkmark & \checkmark   & \checkmark   & \checkmark & ---   \\
\cite{reinhardOptimalCurrentSetpoint2022} & WRSM & \checkmark & $\times$   & $\times$   & $\times$ & $\times$   \\
\cite{parson-scherban2026sdp}             & WRSM & $\times$   & \checkmark & \checkmark & \checkmark & \checkmark \\
\midrule
\rowcolor{proposedgray}
Proposed                                  & WRSM & \checkmark & \checkmark & \checkmark & \checkmark & \checkmark \\
\bottomrule
\multicolumn{7}{l}{\footnotesize $^\dagger$SDP only applicable to WRSM (3-DOF) voltage-constrained regime.}
\end{tabular}
\end{table}

\section{Problem Formulation}
The WRSM can be thought of as a dynamic system which takes a current vector $i = [i_r, i_d, i_q]^\top$ as an input and produces some torque $T_p$ at a certain electrical speed $\omega_e$ as an output. The $i_d$ and $i_q$ elements of the current vector represent the direct-axis and quadrature-axis currents respectively and importantly fully represent the balanced 3-phase AC waveform fed to the motor via the Park-Clarke transform. The WRSM, unlike the PMSM, has one additional current input to control the flux induced by the rotor: $i_r$. The constraints on $i$ are a result of the limitations of the coils through which that current flows. There is also a constraint on the voltage induced in those coils by the spinning magnetic field due to the finite voltage supplying those coils. The strength of the magnetic field $\lambda = [\lambda_r, \lambda_d, \lambda_q]$ induced by the current $i$ is unique for each motor and is represented by an inductance model. 

\label{sec:Prob Form}
\subsection{Local Inductance Model}
For the WRSM, flux linkage and currents are related by the nonlinear map $\lambda = \phi(i)$, $\phi : \mathbb{R}^3 \to \mathbb{R}^3$. Let $\{(i_j, \lambda_j)\}_{j=1}^{N}$ denote samples of $\phi$. In the neighborhood of the $j^{\mathrm{th}}$ current point $i_j$, the nonlinear 
flux map $\phi(i)$ is approximated by the affine map
\begin{align}
    \label{eq:lin approx:eq}
    \phi(i) \approx L_j i + \psi_j,
\end{align}
where $L_j = \nabla \phi(i) \in \mathbb{R}^{3\times 3}$ is the local inductance matrix and $\psi_j = \phi(i_{j}) - \nabla \phi(i)\big|_{i = i_j}i_j$ is the flux offset.\footnote{In order to simplify the notation we use $\nabla$ to represent the Jacobian operator.} $L_r$, $L_d$ and $L_q$ are the rotor, d-axis and q-axis stator self-inductances respectively, $L_{rd}$ is the inductance coupling the rotor to the d-axis, $L_{dq}$ is the cross-coupling inductance and $L_\delta := L_d - L_q$. $L$ is assumed to be symmetric with $L_{rq} = 0$, giving

\begin{align*}
	L = \left [{ \begin{array}{ccc} L_{r} &\quad L_{rd} &\quad 0 \\ L_{rd} &\quad L_{d} &\quad L_{dq}  \\ 0&\quad L_{dq} &\quad L_{q} \\ \end{array} }\right], \quad \psi = \left [{ \begin{array}{c} \psi _{r} \\ \psi _{d} \\ \psi _{q} \\ \end{array} }\right]. 
\end{align*}

The symmetry of $L$ is guaranteed by the physics governing the motor. Cross-saturation, captured by $L_{dq}$, is characterized in \cite{steyaertPiecewiseAffineModeling2023b}. The assumption that the coupling between the rotor and the q-axis is zero is justified by the ordering $L_{rq} < L_{dq} \ll L_d, L_q$, which is typical for WRSMs \cite{eldeebUnifiedTheoryOptimal2018, steyaertPiecewiseAffineModeling2023b}.

\subsection{Optimal Efficiency Reference Generation Problem}
We consider the following QCQP whose solution provides the optimal reference current:
\begin{align*}
    (\mathcal{P})\quad [i^\star, \lambda^\star] &= \argmin_{i,\, \lambda \in \mathbb{R}^3} \quad
    i^\top R i + \omega^2 \lambda^\top G \lambda \tag{O1}\label{eq:P:O1}\\
    \text{s.t.} \quad
    & 0 \leq i_r, \tag{C1}\label{eq:P:C1}\\
    & |i_r| \leq i_{r,\text{r}}, \tag{C2}\label{eq:P:C2}\\
    & \| i_{dq} \| \leq i_{s,\text{r}}, \tag{C3}\label{eq:P:C3}\\
    & |\omega_e| \| \lambda_{dq} \| \leq v_{\text{max}}, \tag{C4}\label{eq:P:C4}\\
    & \tau_p(i, \lambda) = T_p, \tag{C5}\label{eq:P:C5}\\
    & \lambda = L i + \psi. \tag{C6}\label{eq:P:C6}
\end{align*}
The bounds $i_{r,\text{r}}$, $i_{s,\text{r}}$, and $v_{\text{max}}$ denote the rated rotor current, rated stator current magnitude, and maximum phase voltage respectively, and are determined from machine specifications found in Table~\ref{tab:machine_params}. The torque equation $\tau_p(i, \lambda)$ is a nonconvex quadratic as discussed in Section~\ref{sec:torque}. The analysis is presented for $T_p \geq 0$; the case $T_p < 0$ is analogous.

\subsection{Saturation Modeling with PWA Flux Maps} 
\label{sec:PWA_SUB}
Magnetic saturation and cross-saturation are caused by the nonlinear reluctance of iron in the motor \cite{meessenInductanceCalculationsPermanentMagnet2008}. This means the high currents required for high torque and high-speed operation will drive the iron into a region where a local linearization \eqref{eq:lin approx:eq} of the flux map  made at $i = 0$ will result in a large error in the estimate of $\lambda$ \cite{parson-scherbanEffectsMagneticSaturation2024a}. A high torque high current operating point would have smaller inductances $L$ than a local linearization made at $i = 0$. This error in the estimate of $\lambda$ is a weakness of the single-point Taylor approximation inductance map, which is the solution employed in \cite{preindlOptimalStateReference2015a, eldeebUnifiedTheoryOptimal2018, englertOptimalSetpointComputation2018}. This work aims to incorporate magnetic saturation into the motor model by using some other interpolant of $\phi$. Of course, $\phi(i)$ could be interpolated in any number of ways and \cite{steyaertPiecewiseAffineModeling2023b} discusses many standard methods and concludes that PWA provides the most advantages. Most importantly for this work, the PWA map is a continuous interpolant which preserves the quadratic structure of the cost and constraint functions governed by motor physics, and allows the current reference generation problem to be a QCQP within each PWA region. This preservation of the structure allows for the theory of QCQPs in \cite{ConvexOptimizationBoyd, zhouRobustOptimalControl1996} and their SDP relaxations in \cite{barvinokRemarkRankPositive2001a} to be applied.

 The magnetic saturation of the flux map $\phi(i)$ is approximated globally by a PWA function by first partitioning the current space $\mathcal{I} \subset \mathbb{R}^3$ into $N$ simplices via Delaunay triangulation of the Finite Element Analysis (FEA) sample points. Each simplex $\mathcal{I}_j$ is associated with a halfspace representation $\mathcal{I}_j = \{i : A_j i \leq b_j\}$ and an affine flux map $\lambda = L_j i + \psi_j$, where $L_j$ and $\psi_j$ are obtained by least-squares fit to the FEA data within $\mathcal{I}_j$. The result is a continuous PWA approximation of $\phi(i)$ that captures saturation and cross-saturation across the full operating envelope.

\subsection{Loss Model}
\label{sec:loss}
Total electrical loss \eqref{eq:P:O1} is the sum of copper loss and core loss, each a quadratic form. Matrices $R$ ($\Omega$) and $G$ ($1/\Omega$) are fit to FEA data via least squares \cite{steyaertRealTimeCore2023b}. Substituting the local flux map $\lambda = Li + \psi$, the total loss is quadratic in $i$:
\begin{equation*}
    \label{eq:loss}
    \pi(i) = i^\top R_{\mathrm{eff}}\, i + r_{\mathrm{eff}}^\top i + c
\end{equation*}
where $R_{\mathrm{eff}} = R + \omega^2 L^\top G L \succeq 0$, $r_{\mathrm{eff}} = 2\omega^2 L^\top G \psi$, $c = \omega^2 \psi^\top G \psi$. The constant $c$ does not affect the minimizer and is neglected.

\subsection{Field Weakening Model}
Substituting \eqref{eq:P:C6} into \eqref{eq:P:C4} yields the quadratic voltage constraint
\begin{equation*}
    \label{eq:voltage:quad}
    i^\top Q_v\, i + q_v^\top i + c_v \leq 0,
\end{equation*}
where
\begin{align*}
    Q_v &= \begin{bmatrix} L_{rd}^2 & L_{rd}L_d & L_{rd}L_{dq} \\ L_{rd}L_d & L_d^2 + L_{dq}^2 & L_dL_{dq} + L_{dq}L_q \\ L_{rd}L_{dq} & L_dL_{dq} + L_{dq}L_q & L_{dq}^2 + L_q^2 \end{bmatrix}, \\
    q_v &= 2\begin{bmatrix} L_{rd} & 0 \\ L_d & L_{dq} \\ L_{dq} & L_q \end{bmatrix} \begin{bmatrix} \psi_d \\ \psi_q \end{bmatrix}, \quad
    c_v = \left\| \begin{bmatrix} \psi_d \\ \psi_q \end{bmatrix} \right\|^2 - \bar{v}^2,
\end{align*}
and $\bar{v} = v_{\max}/|\omega_e|$. As $Q_v \succeq 0$, so the constraint is convex.

\subsection{Torque Model}
\label{sec:torque}
The torque per pole pair \eqref{eq:P:C5} is $\tau_p(i,\lambda) = i^\top J \lambda$ where
\begin{equation*}
    J = \begin{bmatrix} 0&0&0\\0&0&-1\\0&1&0 \end{bmatrix}.
\end{equation*}
Substituting \eqref{eq:P:C6} into \eqref{eq:P:C5} yields the quadratic torque constraint
\begin{equation*}
    \label{eq:torque:quad}
    i^\top Q_\tau\, i + q_\tau^\top i = T_p,
\end{equation*}
where $Q_\tau = \tfrac{1}{2}(JL + L^\top J^\top)$ and $q_\tau = J\psi$:
\begin{equation*}
    Q_\tau =
    \begin{bmatrix}
        0 & 0 & \frac{L_{rd}}{2} \\[4pt]
        0 & -L_{dq} & \frac{L_\delta}{2} \\[4pt]
        \frac{L_{rd}}{2} & \frac{L_\delta}{2} & L_{dq}
    \end{bmatrix}, \quad
    q_\tau =
    \begin{bmatrix}
        0 \\ -\psi_q \\ \psi_d
    \end{bmatrix}.
\end{equation*}
The torque equality constraint is a quadratic equality and is therefore nonconvex. Since $\operatorname{tr}(Q_\tau) = 0$ and $Q_\tau \neq 0$, its eigenvalues sum to zero and are not all zero, so $Q_\tau$ is indefinite. 

\section{Constraint Activation Regimes and Closed Form Solutions}
\label{sec:Regime+Solution}
The optimal current solution $i^\star$ from ($\mathcal P$) depends entirely on which constraints bind at a given operating point. Each regime corresponds to 
a characteristic driving behavior: low-demand steady operation 
(cruise), high-torque launch from rest 
(launch control), high-speed operation within the voltage limit (fast driving), and simultaneous 
current and voltage saturation under aggressive high-speed 
operation (forceful fast driving). The rotor 
current-constrained regime, when either \eqref{eq:P:C1} or \eqref{eq:P:C2} is binding, recovers the two degree-of-freedom case 
of~\cite{eldeebUnifiedTheoryOptimal2018} as a special case.
The active constraint structure thus admits direct interpretation 
as driving behavior, connecting the mathematical regime 
classification to physically meaningful operating conditions 
encountered in a representative drive cycle. 

\subsection{Torque-Only Regime: Cruise}
When all of the inequality constraints are slack, ($\mathcal{P}$) reduces to:
\begin{equation}
    \label{eq:cruise}
    \min_{i \in \mathbb{R}^3} \quad i^\top R_{\mathrm{eff}}\, i + r_{\mathrm{eff}}^\top i
    \qquad \text{s.t.} \quad i^\top Q_\tau\, i + q_\tau^\top i = T_p.
\end{equation}
The KKT stationarity condition \cite{zhouRobustOptimalControl1996} yields the $3\times 3$ linear system
\begin{equation*}
    A(\mu)\,i = -b(\mu), \qquad b(\mu) := r_{\mathrm{eff}} + \mu q_\tau,
\end{equation*}
where $\mu$ is the Lagrange multiplier, $A(\mu) := 2(R_{\mathrm{eff}} + \mu Q_\tau)$
is affine in $\mu$ with entries determined by the machine resistance, inductance,
and core-loss parameters, and $b(\mu)$ is affine in $\mu$ with entries determined
by the core-loss parameters and flux offset $\psi$. The optimal current is
\begin{equation*}
    i^\star = i(\mu^\star), \qquad
    i(\mu) = -A(\mu)^{-1}b(\mu),
\end{equation*}
where $\mu^\star$ is the real root of $\sum_{k=0}^{6}a_k\mu^k=0$
(see Appendix~\ref{app:caseA}) minimizing
$i(\mu)^\top R_{\mathrm{eff}}\,i(\mu) + r_{\mathrm{eff}}^\top i(\mu)$.

\subsection{Stator Current Constrained Regime: Launch Control}
When~\eqref{eq:P:C3} is binding and \eqref{eq:P:C1}, \eqref{eq:P:C2}, \eqref{eq:P:C4}, are slack,
$(\mathcal{P})$ reduces to:
\begin{equation}
    \label{eq:Launch}
    \begin{aligned}
        \min_{i \in \mathbb{R}^3} \quad & i^\top R_{\mathrm{eff}}\, i + r_{\mathrm{eff}}^\top i \\
        \text{s.t.} \quad
        & i^\top Q_\tau\, i + q_\tau^\top i = T_p, \\
        & i_d^2 + i_q^2 = i_{s,\mathrm{r}}^2.
    \end{aligned}
\end{equation}
Parametrizing the stator circle as
$i_d = i_{s,\mathrm{r}}\cos\theta$, $i_q = i_{s,\mathrm{r}}\sin\theta$
and noting that the $(1,1)$ entry of $Q_\tau$ is zero,
substituting into the torque constraint uniquely determines
\begin{align*}
    i_r(\theta) &= \frac{T_p - \gamma(\theta)}{L_{rd}\,i_{s,\mathrm{r}}\sin\theta}, \\
    \gamma(\theta) &= i_{s,\mathrm{r}}^2\bigl(L_{dq}(\sin^2\theta-\cos^2\theta)
    + L_\delta\cos\theta\sin\theta\bigr) \\
    &\quad+ i_{s,\mathrm{r}}(\psi_d\sin\theta-\psi_q\cos\theta),
\end{align*}
reducing~\eqref{eq:Launch} to univariate optimization in~$\theta$.
Note that $\sin\theta \to 0$ implies $i_r \to \infty$, violating the rotor
current limit; such $\theta$ would be handled by the PMSM case in Section \ref{sec:PMSM}.
The stationarity condition, after clearing denominators, yields a
trigonometric polynomial of degree at most~$5$ in $(\cos\theta,\sin\theta)$.
The Weierstrass substitution $t = \tan(\theta/2)$, with
$\cos\theta = (1-t^2)/(1+t^2)$ and $\sin\theta = 2t/(1+t^2)$,
converts this to $\sum_{k=0}^{10}a_k t^k = 0$
(coefficients omitted for brevity). The optimal current is
\begin{equation*}
    i^\star = i(t^\star), \quad
    i(t) = \bigl[i_r(\theta(t)),\; i_{s,\mathrm{r}}\cos\theta(t),\; i_{s,\mathrm{r}}\sin\theta(t)\bigr]^\top,
\end{equation*}
where $\theta(t) = 2\arctan(t)$, and $t^\star$ is the real root minimizing
$i(t)^\top R_{\mathrm{eff}}\,i(t) + r_{\mathrm{eff}}^\top i(t)$,
found via companion matrix eigendecomposition.

\subsection{Voltage-Constrained Regime: Fast Driving}
When~\eqref{eq:P:C4} is binding and \eqref{eq:P:C1}, \eqref{eq:P:C2}, \eqref{eq:P:C3}, are slack,
$(\mathcal{P})$ reduces to:
\begin{equation}
    \label{eq:Fast}
    \begin{aligned}
        \min_{i \in \mathbb{R}^3} \quad & i^\top R_{\mathrm{eff}}\, i + r_{\mathrm{eff}}^\top i \\
        \text{s.t.} \quad
        & i^\top Q_\tau\, i + q_\tau^\top i = T_p, \\
        & i^\top Q_v\, i + q_v^\top i = \bar{v}^2.
    \end{aligned}
\end{equation}
This is an inhomogeneous QCQP with two quadratic equality constraints
in~$\mathbb{R}^3$. Homogenizing via $\tilde{i} = [i^\top, 1]^\top$ and lifting
$X = \tilde{i}\tilde{i}^\top$ yields the SDP relaxation
\begin{equation}
    \label{eq:SDR}
    \begin{aligned}
        \min_{X \succeq 0} \quad & \tr(\tilde{R}_{\mathrm{eff}}\, X) \\
        \text{s.t.} \quad
        & \tr(\tilde{Q}_\tau\, X) = T_p, \\
        & \tr(\tilde{Q}_v\, X) = \bar{v}^2, \\
        & X_{4,4} = 1, 
    \end{aligned}
\end{equation}
where $\tilde{R}_{\mathrm{eff}}$, $\tilde{Q}_\tau$,
$\tilde{Q}_v \in \mathbb{S}^4$ are the homogenized forms of the
objective, torque, and voltage matrices, respectively.
Since $R_{\mathrm{eff}} \succ 0$, the dual is strictly feasible~\cite{ConvexOptimizationBoyd}.
Rank-1 optimality of the SDP solution, which certifies that the relaxation
is exact and recovers a global minimizer of~\eqref{eq:Fast}, is
verified computationally across the operating envelope. This approach was applied to a similar problem in  \cite{parson-scherban2026sdp} with optimality guarantees coming from \cite{barvinokRemarkRankPositive2001a}.

\subsection{Fully Constrained Regime: Forceful Fast Driving}
When~\eqref{eq:P:C3}, \eqref{eq:P:C4}, are binding and  \eqref{eq:P:C1}, \eqref{eq:P:C2} are slack,
$(\mathcal{P})$ reduces to:
\begin{equation}
    \label{eq:forceful fast}
    \begin{aligned}
        \min_{i \in \mathbb{R}^3} \quad & i^\top R_{\mathrm{eff}}\, i + r_{\mathrm{eff}}^\top i \\
        \text{s.t.} \quad
        & i^\top Q_\tau\, i + q_\tau^\top i = T_p, \\
        & i^\top Q_v\, i + q_v^\top i = \bar{v}^2, \\
        & i_d^2 + i_q^2 = i_{s,\mathrm{r}}^2.
    \end{aligned}
\end{equation}
The feasible set is the intersection of three quadric surfaces in $\mathbb{R}^3$ ---
generically a finite set of points --- so the problem reduces from constrained
optimization to algebraic root-finding. Parametrizing the stator circle as
$i_d = i_{s,\mathrm{r}}\cos\theta$, $i_q = i_{s,\mathrm{r}}\sin\theta$,
for fixed $\theta$ both constraints are quadratic in $i_r$:
\begin{align*}
    a_1(\theta)\,i_r^2 + b_1(\theta)\,i_r + c_1(\theta) &= 0, \qquad \text{(torque)} \\
    a_2(\theta)\,i_r^2 + b_2(\theta)\,i_r + c_2(\theta) &= 0, \qquad \text{(voltage)}
\end{align*}
where $a_j, b_j, c_j$ are trigonometric polynomials in $\theta$.
Eliminating $i_r$ via the resultant of the two quadratics yields a single trigonometric
equation in $\theta$ alone. The Weierstrass substitution $t = \tan(\theta/2)$ converts
this to a univariate polynomial in $t$ of degree at most 4, whose roots are found via
companion matrix eigendecomposition. Coefficients of that polynomial are provided in Appendix~\ref{app:caseC}. For each real root $t_k$, the corresponding
$(i_d, i_q)$ are recovered and $i_r$ is obtained. The candidate with minimum loss is the solution for this regime, as the procedure exhaustively enumerates all constraint-satisfying points.

\subsection{Rotor Current Constrained Regime: PMSM Case}
\label{sec:PMSM}
When~\eqref{eq:P:C1} or~\eqref{eq:P:C2} are binding, $i_r$ is fixed at its bound.
Defining $\psi_d^{\mathrm{eff}} := L_{rd}i_r + \psi_d$,
($\mathcal{P}$) reduces to a two-dimensional QCQP in~$(i_d, i_q)$
structurally identical to the PMSM optimal reference generation
problem solved in \cite{preindlOptimalStateReference2015a} for the case without cross-saturation ie $L_{dq} = 0$. Closed-form solutions with nonzero $L_{dq}$ for all sub-cases: only (\ref{eq:P:C3}) binding, only (\ref{eq:P:C4}) binding, and both binding
follow directly
from~\cite{eldeebUnifiedTheoryOptimal2018} with
$\psi_d^{\mathrm{eff}}$ replacing~$\psi_{\mathrm{pm}}$.

\section{Results}

\subsection{Experimental Setup}
\begin{figure}[t]
    \centering
    \vspace{8pt}
    \fbox{\includegraphics[width=.6\columnwidth]{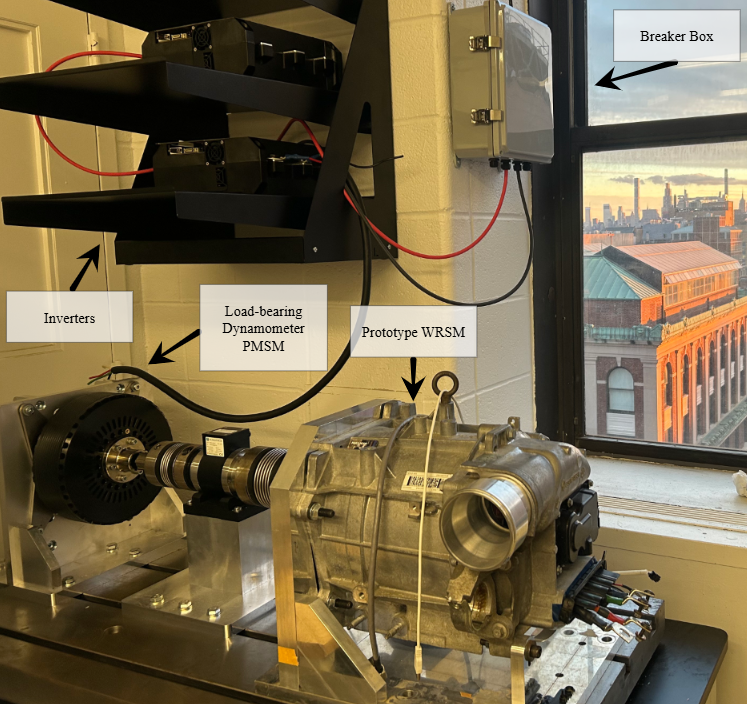}}
    \caption{Prototype WRSM.}
    \label{fig:motor}
    \vspace{-10pt}
\end{figure}
The proposed framework is evaluated on a prototype WRSM whose nameplate
parameters are listed in Table~\ref{tab:machine_params}. The nonlinear flux
map $\phi(i)$ was obtained via FEA of the machine
geometry and materials. The current space $\mathcal{I} \subset \mathbb{R}^3$
was sampled at a uniform grid of FEA operating points and partitioned into
simplices via Delaunay triangulation, yielding a continuous PWA approximation
of $\phi(i)$ as described in Section~\ref{sec:PWA_SUB}. The loss matrices
are $R = \mathrm{diag}(0.004,\,0.045,\,0.045)$ in ($\Omega$) and
$G = \mathrm{diag}(0,\,0.0033,\,0.0092)$ in ($1/\Omega$), fit to FEA data as
in~\cite{steyaertRealTimeCore2023b}. Representative inductance values and
flux offsets at nameplate and high-torque operating points are reported in
Table~\ref{tab:flux_approx}, illustrating the degree of saturation present
at high current: the $d$-axis inductance $L_d$ drops by an factor of seven
between no-load and high torque, confirming that a single-point linearization
would introduce substantial error across the operating envelope. All timing
experiments were conducted on a 12th Generation Intel Core i7-1255U. The
prototype machine is shown in Fig.~\ref{fig:motor}.

\begin{table}[t]
    \renewcommand{\arraystretch}{1.1}
    \setlength{\tabcolsep}{4pt}
    \vspace{8pt}
    \centering
    \caption{Prototype WRSM Parameters}
    \label{tab:machine_params}
    \footnotesize
    \begin{tabular}{l r l r}
        \toprule
        \textbf{Parameter} & \textbf{Value} &
        \textbf{Parameter} & \textbf{Value} \\
        \midrule
        Turns ratio $N_f/N_s$ & 39 & Pole pairs $p$ & 2 \\
        Stator resistance $R_s$ [m$\Omega$] & 45 &
        Rotor resistance $R_r$ [m$\Omega$] & 4 \\
        Max torque [Nm] & 130 & Base speed [rpm] & 3000 \\
        Max speed [rpm] & 12000 & DC-link voltage [V] & 325 \\
        Max power [kW] & 65 & & \\
        \bottomrule
    \end{tabular}
\end{table}

\begin{table}[!t]
    \renewcommand{\arraystretch}{1.1}
    \setlength{\tabcolsep}{4pt}
    \centering
    \caption{Flux approximations at nameplate and peak operating points.}
    \label{tab:flux_approx}
    \footnotesize
    \begin{tabular}{lcc|lcc}
        \toprule
        \textbf{Param.} & \textbf{0\,Nm} & \textbf{80\,Nm} & \textbf{Param.} & \textbf{0\,Nm} & \textbf{80\,Nm} \\
        \midrule
        $i_r$ [A]        & 0   & 66.9     & $L_r$    [mH] & 2.1 & 0.033    \\
        $i_d$ [A]        & 0   & $-$39.1  & $L_d$    [mH] & 2.4 & 0.331    \\
        $i_q$ [A]        & 0   & 173.5    & $L_q$    [mH] & 0.8 & 0.257    \\
        $\psi_r$ [Wb]    & 0   & 0.249    & $L_{rd}$ [mH] & 2.1 & 0.053    \\
        $\psi_d$ [Wb]    & 0   & 0.229    & $L_{dq}$ [mH] & 0   & $-$0.067 \\
        $\psi_q$ [Wb]    & 0   & 0.058    &               &     &          \\
        \bottomrule
    \end{tabular}
\end{table}

\subsection{Constraint Regime Classification}

\begin{figure}[t]
    \centering
    \vspace{10pt}
    \includegraphics[width=.5\columnwidth]{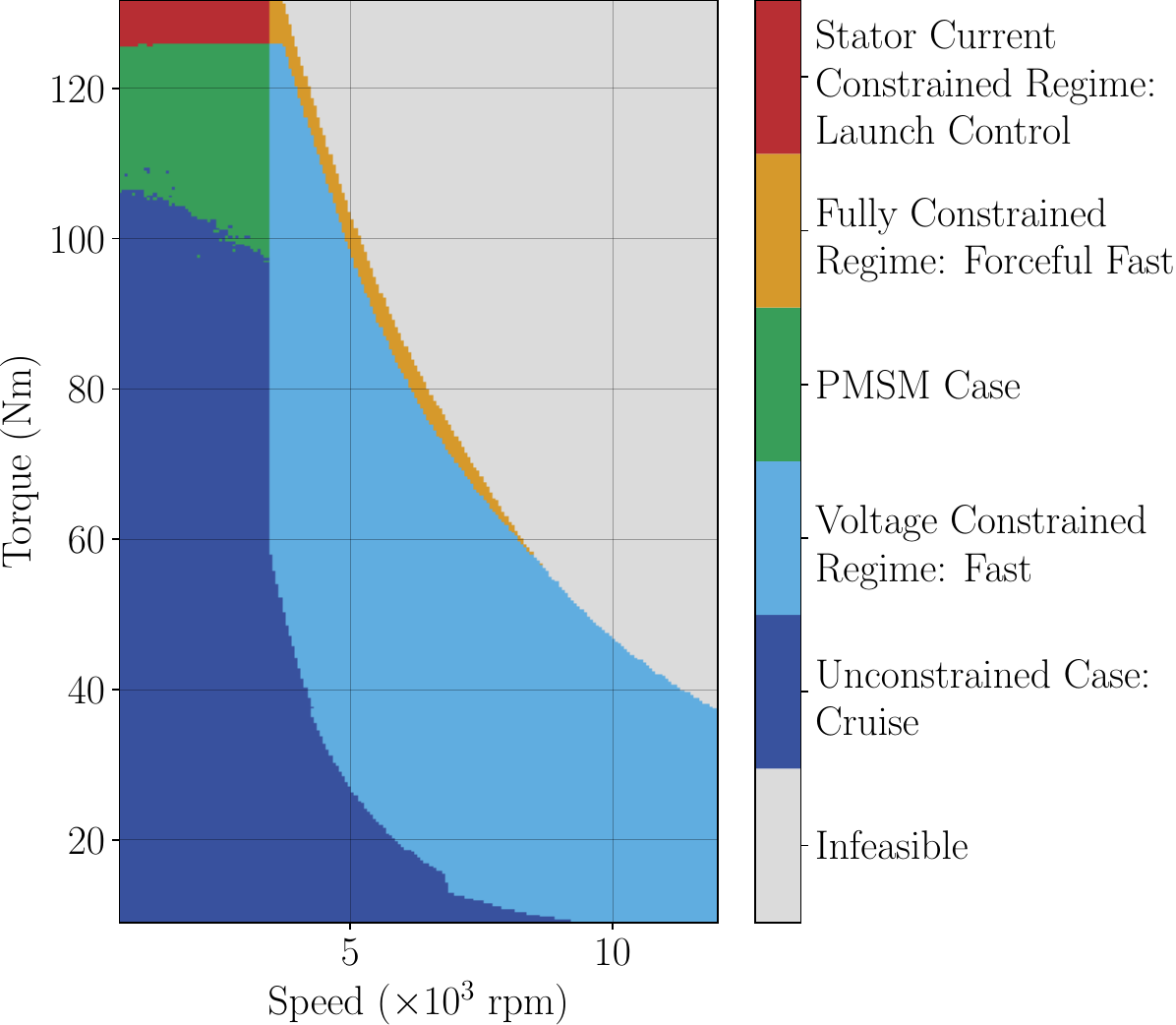}
    \caption{Constraint regimes across the torque-speed envelope of the
             prototype WRSM. The torque-dependent cruise-to-voltage boundary
             near 5000\,rpm is characteristic of the WRSM's active flux
             control capability.}
    \label{fig:regimes_pwa}
    \vspace{-10pt}
\end{figure}

The five constraint regimes identified in Section~\ref{sec:Regime+Solution}
partition the torque-speed operating envelope in a physically interpretable
way, as shown in Fig.~\ref{fig:regimes_pwa}. At low speed and moderate
torque the unconstrained cruise regime dominates, where the optimal solution
is determined solely by the torque equality and the loss objective. As torque
increases the stator current limit activates, yielding the launch-control
regime. Above the base speed of 3000\,rpm the voltage constraint engages,
transitioning operation into the fast-driving and, at high torque and speed,
the forceful-fast-driving regime. At high speeds and low torque the rotor
current saturates, recovering the two-degree-of-freedom PMSM subproblem.

A feature distinctive to the WRSM is visible near 5000\,rpm: the boundary
between the cruise and voltage-constrained regimes is not a fixed speed
threshold but varies with the requested torque. This torque-dependence arises
because the rotor current degree of freedom allows the machine to modulate
its air-gap flux, shifting the speed at which the voltage constraint first
binds. No analogous behavior exists in the PMSM, where the flux is fixed by
the permanent magnet. This confirms that the three degree-of-freedom
formulation captures operating characteristics that a two degree-of-freedom
model cannot represent.

\subsection{Optimality Validation}

\begin{figure}[!h]
    \centering
    \begin{subfigure}[t]{0.48\linewidth}
        \centering
        \includegraphics[width=\linewidth]{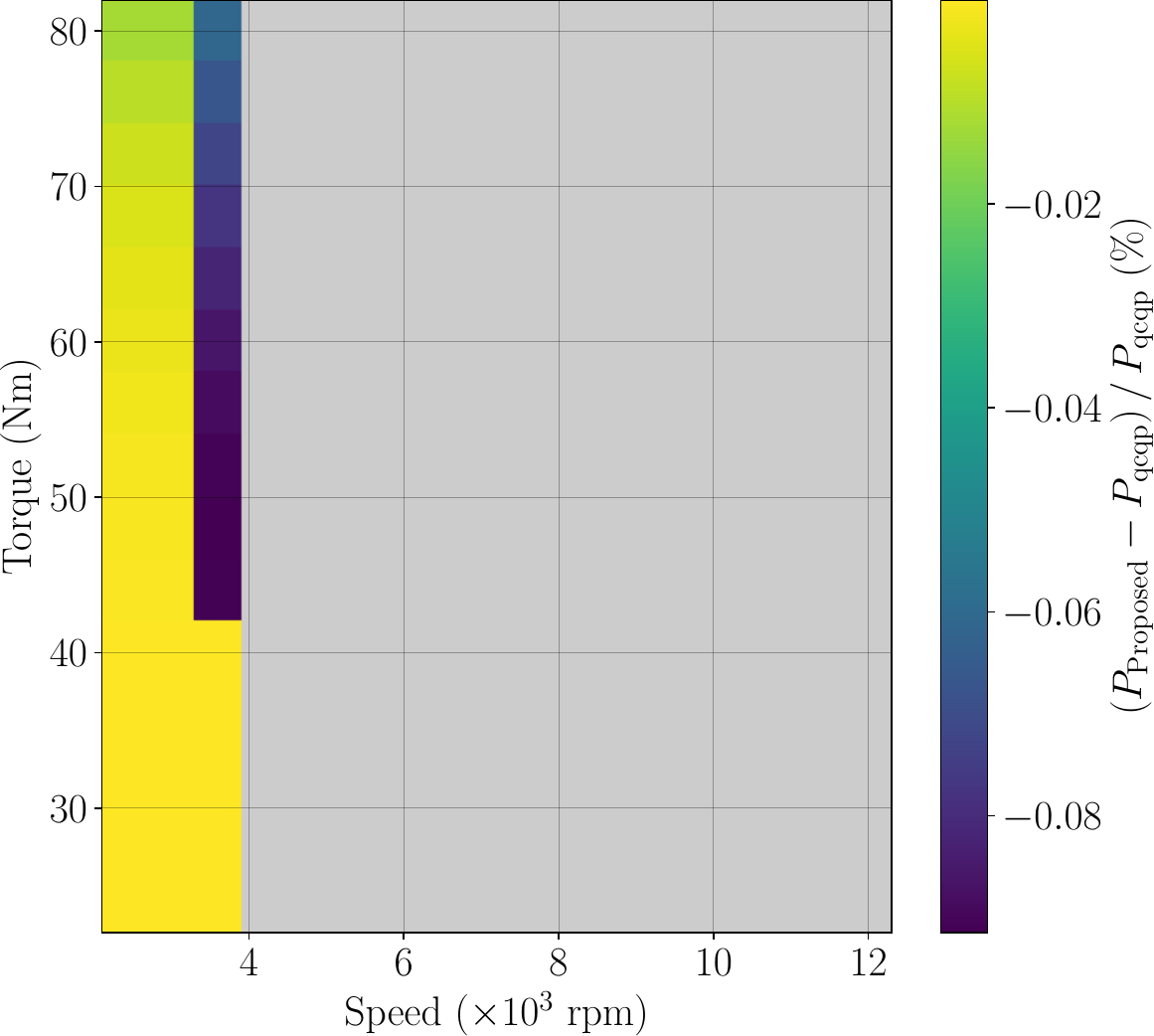}
        \caption{Pointwise cost difference: the proposed method finds points with a lower objective than Gurobi. The difference is below 1\,W throughout the tested envelope.}
        \label{fig:loss_diff}
    \end{subfigure}
    \hfill
    \begin{subfigure}[t]{0.5\linewidth}
        \centering
        \includegraphics[width=\linewidth]{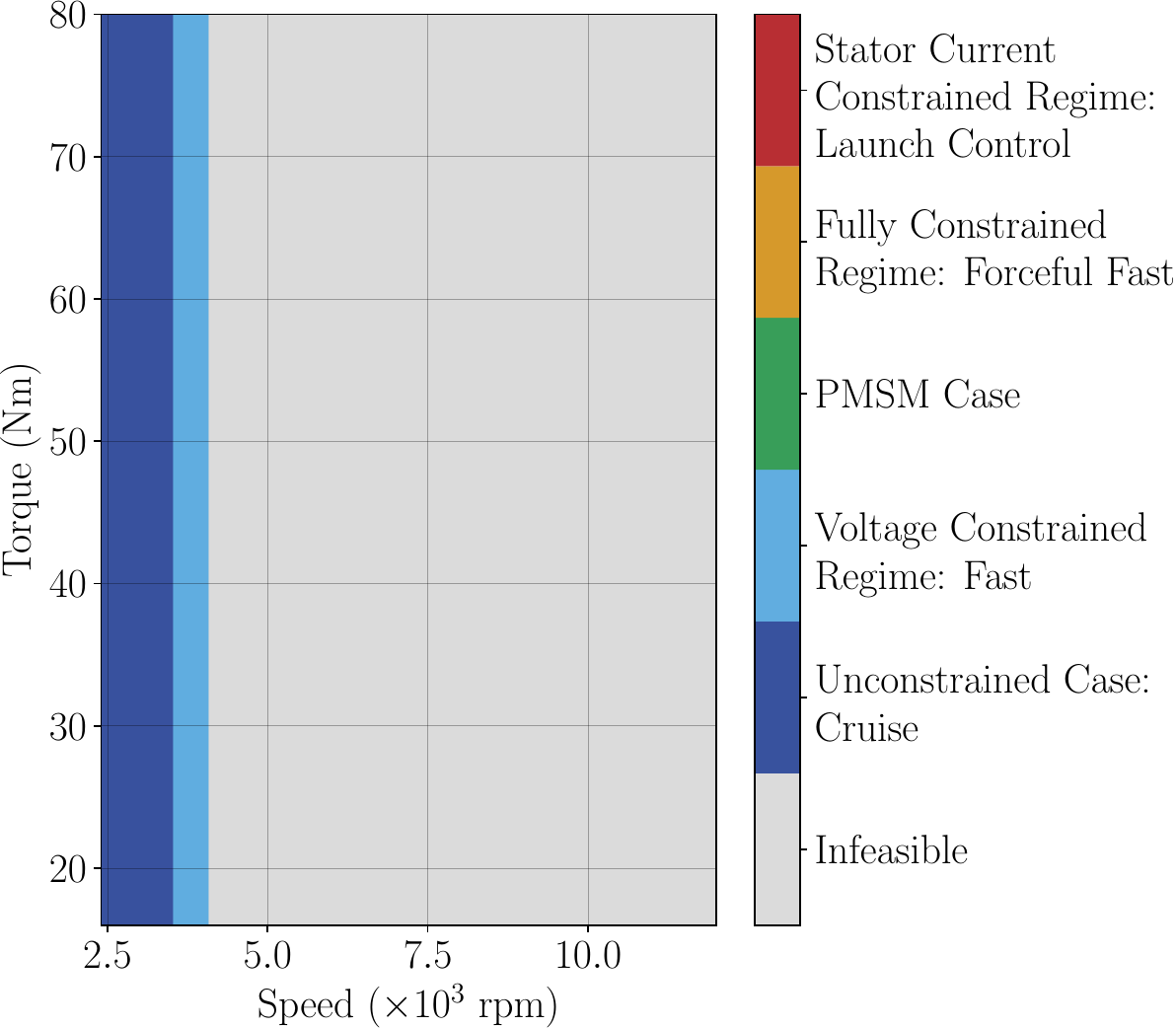}
        \caption{Constraint regimes active over the candidate PWA region.}
        \label{fig:regimes_qcqp}
    \end{subfigure}
    \caption{Per-tile optimality validation and regime classification.}
    \label{fig:optimality}
\end{figure}

The proposed framework is validated by comparing its solutions to those
returned by the state-of-the-art nonlinear QCQP solver
Gurobi~\cite{gurobi} across $n = 101$ operating points in a candidate PWA
region. Fig.~\ref{fig:loss_diff} shows the pointwise difference in optimal
cost: the proposed method improves upon Gurobi up to 1\,W across the entire
tested envelope, which is an improvement upon the conventional methods. The regime classification for the same candidate region
is shown in Fig.~\ref{fig:regimes_qcqp}, confirming that the closed-form
cruise solution and the SDP-based fast-driving solution are each applied
in the correct portion of the operating space.

\begin{figure}[!h]
    \centering
    \begin{subfigure}[t]{0.48\linewidth}
        \centering
        \vspace{6pt}
        \includegraphics[width=.9\linewidth]{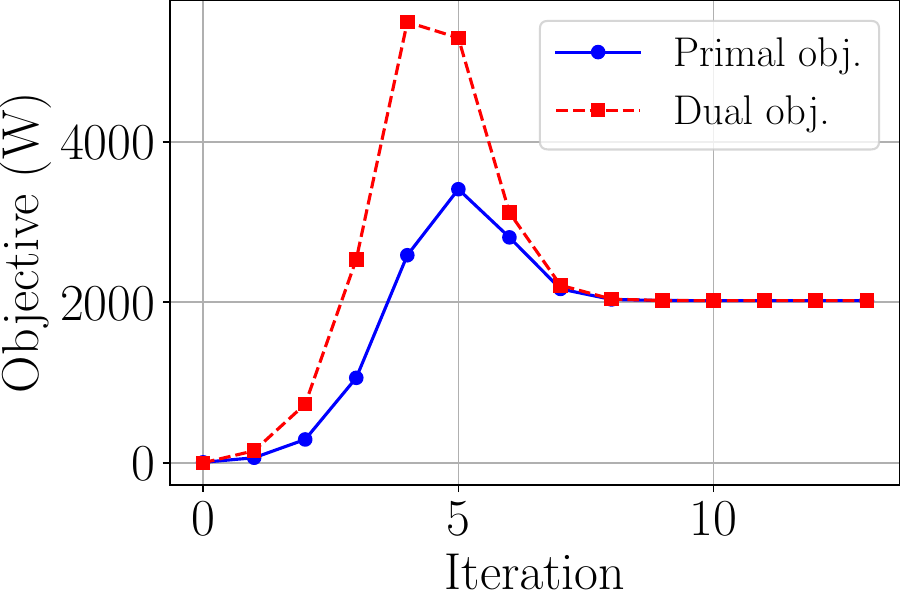}
        \caption{Cost convergence.}
        \label{fig:mosek_cost}
    \end{subfigure}
    \hfill
    \begin{subfigure}[t]{0.48\linewidth}
        \centering
        \vspace{6pt}
        \includegraphics[width=.9\linewidth]{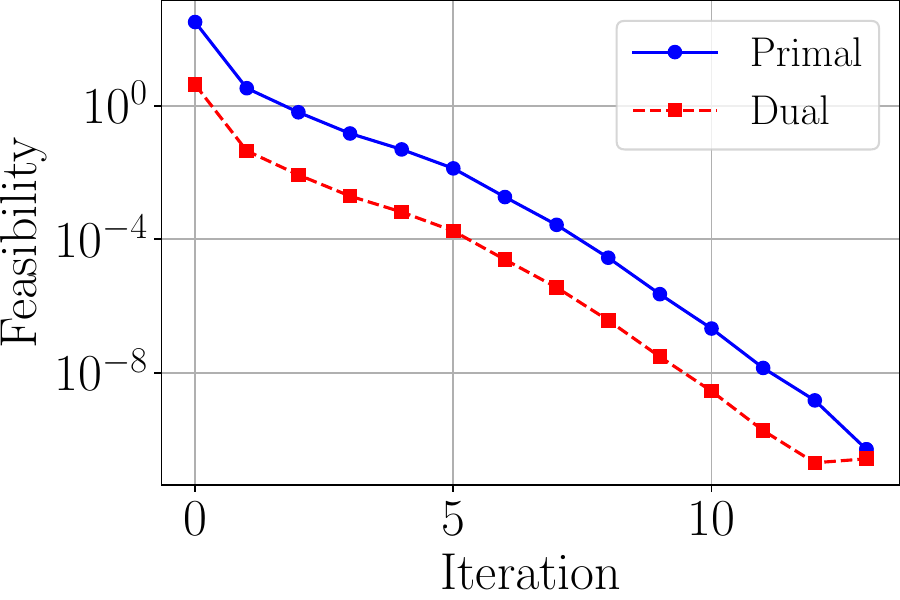}
        \caption{Feasibility convergence.}
        \label{fig:mosek_feas}
    \end{subfigure}
    \caption{MOSEK convergence for a representative fast-driving operating
             point ($T_p = 20$\,Nm, $\omega_e = 500$\,rad/s). Primal-dual
             convergence is reached in 13 iterations.}
    \label{fig:mosek}
\end{figure}

Optimality guarantees differ by regime. In the cruise, launch-control,
forceful-fast-driving, and rotor-constrained regimes, global optimality
follows from the KKT conditions: the closed-form candidate solutions are
derived by satisfying stationarity exactly, and the feasible candidate with
minimum cost is selected by exhaustive enumeration over the finite root set.
In the voltage-constrained (fast-driving) regime, the SDP
relaxation~\eqref{eq:SDR} is solved via MOSEK~\cite{mosek}. Global optimality is certified by verifying
that the returned solution $X^\star$ has rank one: a rank-1 solution certifies
that the relaxation is tight and that a globally optimal primal solution has
been recovered~\cite{barvinokRemarkRankPositive2001a}. Rank-1 recovery was
confirmed at every tested operating point in the voltage-constrained regime.
Fig.~\ref{fig:mosek} illustrates the cost and feasibility convergence of a
representative MOSEK solve at $T_p = 20$\,Nm, $\omega_e = 500$\,rad/s,
reaching primal-dual convergence in 13 iterations, which is typical for
SDPs of this size ($X \in \mathbb{S}^4$).

\begin{figure*}[t]
    \centering
    \vspace{5pt}
    \includegraphics[width=.9\textwidth]{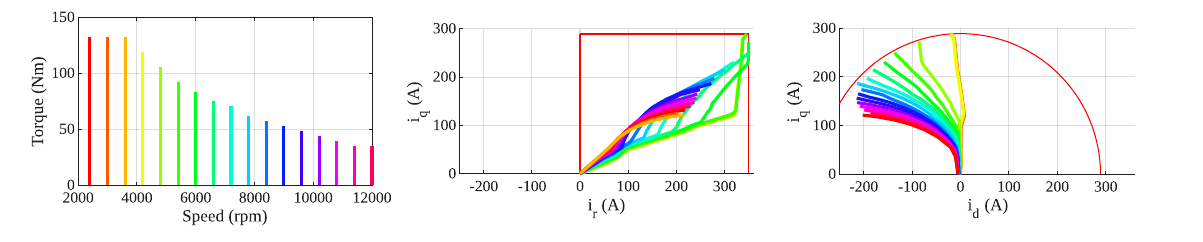}
    \caption{Optimal current trajectories across the operating range of the
             prototype WRSM. Color encodes electrical speed. The rotor current
             $i_r$ decreases at high speed due to voltage-constraint-driven
             field weakening. The shift toward negative $i_d$ with speed is
             consistent with flux reduction in the field-weakening regime.}
    \label{fig:traj}
\end{figure*}

Fig.~\ref{fig:traj} shows the optimal current trajectories across the full
torque-speed operating range, with color indicating electrical speed. Several
physically meaningful trends are visible. In the $i_r$--$i_d$ plane, the
rotor current decreases at high speed as the voltage constraint forces the
machine into field weakening: reducing $i_r$ lowers the air-gap flux and
relaxes the voltage limit, at the cost of requiring larger stator currents to
maintain torque. In the $i_d$--$i_q$ plane, the trajectories shift toward
more negative $i_d$ with increasing speed, consistent with field-weakening
operation. As speed increases, the bending of paths within the same regime reflects the increasing influence of the core-loss penalty
$\omega_e^2 \lambda^\top G \lambda$ at high electrical frequencies.

\subsection{Runtime Comparison}

The proposed framework achieves a \textbf{95.5\% reduction in median solve
time} relative to Gurobi, with a median per-tile solve time of 0.99\,ms
compared to 22.22\,ms for the QCQP baseline (Table~\ref{tab:solver_comparison}).
This reduction is a direct consequence of the regime-aware solution strategy:
in all the active regimes from among cruise, launch-control, fast-driving, forceful-fast-driving, and rotor-constrained regimes, the globally optimal solution is obtained via closed-form expressions
or companion-matrix eigendecomposition, with no iterative solver invoked. The
SDP solver is called only in the voltage-constrained regime, which corresponds
to a subset of the high-speed operating region (Fig.~\ref{fig:regimes_pwa}).

The worst-case MOSEK time of 4.14\,ms occurs within this regime and represents
the computational ceiling of the proposed approach. Even in the worst case,
the proposed framework is more than five times faster than the Gurobi median.
The median solve time of 0.99\,ms promises future satisfaction of real-time deployment
requirements in embedded motor drive hardware, where reference generation
must complete within a single control period typically on the order of
10--50$\mu$s at standard switching frequencies.

\begin{table}[!h]
\renewcommand{\arraystretch}{1.1}
\centering
\caption{Runtime comparison: proposed framework vs.\ Gurobi QCQP baseline.}
\label{tab:solver_comparison}
\begin{tabular}{llccc}
    \toprule
    Solver & Method & Median [ms] & Max [ms] & Mean [ms] \\
    \midrule
    Gurobi        & QCQP             & 22.22 & 23.49 & 18.12 \\
    Proposed      & Regime-aware     &  0.99 &  4.14 &  1.09 \\
    \midrule
    \multicolumn{2}{l}{Reduction}    & 95.5\% & 82.4\% & 94.0\% \\
    \bottomrule
\end{tabular}
\end{table}

\section{Conclusion}
\label{sec:conclusion:sec}
This work generalizes unified optimal current reference theory to 
three-degree-of-freedom wound rotor synchronous machines, 
incorporating magnetic saturation, cross-coupling, and 
speed-dependent core losses via a piecewise-affine flux map. 
Closed-form KKT solutions are derived for the cruise, launch 
control, fast driving, and forceful fast driving regimes, and a verifiably exact SDP 
relaxation handles the voltage-constrained fast driving regime. 
Validation on a physical WRSM prototype demonstrates that the 
proposed framework outperforms a state-of-the-art nonlinear solver 
by up to 1\,W while achieving a 95\% reduction in median solve 
time, with a median per-tile solve time under 1\,ms. Future work 
includes formal exactness certification and finding optimality bounds of the SDP relaxation 
via rank conditions and extension to online implementation in 
embedded motor drive hardware.

\section*{Acknowledgment}
This work was sponsored by Tau Motors.
\bibliographystyle{IEEEtran}
\bibliography{references}

\appendix
\section{Polynomial Coefficients for Torque-Only Regime: Cruise}
\label{app:caseA}
In the cruise regime the problem reduces to~\eqref{eq:cruise}. KKT stationarity gives $i(\mu) = -A(\mu)^{-1}b(\mu)$ with $A(\mu):=2(R_\mathrm{eff}+\mu Q_\tau)$ and $b(\mu):=[b_r,\; b_d-\mu\psi_q,\; b_q+\mu\psi_d]^\top$. Substituting $-A(\mu)^{-1}b(\mu)$ into the torque constraint yields $\sum_{k=0}^{6}a_k\mu^k=0$ with
\begin{align*}
a_6 &= -L_{dq}L_{rd}^4\zeta, \\
a_5 &= 2L_{rd}^2\kappa\zeta, \\
a_4 &= -4\kappa^2 T_p - L_{dq}L_{rd}^2(L_\delta^2+4L_{dq}^2)b_r^2 + 2L_{rd}(\alpha^\top\psi+\beta^\top b)b_r \\
    &\quad - 4R_r^2 L_{dq}(L_\delta^2+4L_{dq}^2)\psi^\top D\psi + b^\top M_1\psi + b^\top M_2 b + \psi^\top M_3\psi, \\
a_3 &= -8L_{rd}^2\bigl[- L_{dq}^2 R_r b^\top D b + L_\delta L_{dq}R_r b_d b_q + L_{dq}L_{rd}R_s b_q b_r \\
    &\quad + 4L_{dq}R_r R_s^2 T_p + 2L_{dq}R_r R_s b^\top J\psi - 2R_r R_s^2\psi_q^2\bigr], \\
a_2 &= 32R_r R_s^2\kappa T_p + 12L_{dq}L_{rd}^2 R_s^2 b_r^2 + \bigl[(4R_r R_s L_{rd}(L_\delta^2+4L_{dq}^2) + 4R_s^2 L_{rd}^3)b_q \\
    &\quad + 24R_r R_s^2 L_{rd}(L_\delta\psi_q - 2L_{dq}\psi_d)\bigr]b_r + 4R_r^2 L_\delta(L_\delta^2+4L_{dq}^2)b_d b_q \\
    &\quad - 4R_r^2 L_{dq}(L_\delta^2+4L_{dq}^2) b^\top D b + 8R_r^2 R_s(L_\delta^2+4L_{dq}^2) b^\top J\psi \\
    &\quad + b^\top M_4 b + b^\top M_5\psi + \psi^\top M_6\psi, \\
a_1 &= -8R_s\bigl[(L_\delta^2+4L_{dq}^2)R_r^2|b|^2 + L_{rd}^2 R_s(R_s b_r^2+R_r b_q^2) + 4R_r^2 R_s^2|\psi|^2 \\
    &\quad - 4R_r^2 R_s(L_\delta\psi^\top D b + 2L_{dq}(b_d\psi_q+b_q\psi_d)) \\
    &\quad + L_{rd}R_r R_s b_r(2L_\delta b_d + 4L_{dq}b_q - 4R_s\psi_d)\bigr], \\
a_0 &= 16R_r R_s^2\bigl[- L_{dq}R_r b^\top D b + L_\delta R_r b_d b_q + L_{rd}R_s b_q b_r - 4R_r R_s^2 T_p + 2R_r R_s b^\top J\psi\bigr],
\end{align*}
where $\kappa:=R_r(L_\delta^2+4L_{dq}^2)+L_{rd}^2 R_s$, $\zeta:=4L_{dq}T_p-\psi_q^2$, and
\begin{align*}
\alpha &= R_r\begin{bmatrix} 2L_\delta^2 L_{dq}+8L_{dq}^3\\ -(L_\delta^3+4L_\delta L_{dq}^2) \end{bmatrix} +R_s\begin{bmatrix}0\\-L_\delta L_{rd}^2\end{bmatrix}, \quad
\beta = \begin{bmatrix} L_\delta L_{dq}L_{rd}^2\\2L_{dq}^2 L_{rd}^2 \end{bmatrix}, \\
D &:= \begin{bmatrix}1&0\\0&-1\end{bmatrix}, \quad J := \begin{bmatrix}0&1\\-1&0\end{bmatrix}.
\end{align*}
\begin{align*}
M_1 &= L_{rd}^2\Bigl(R_r\begin{bmatrix}-4L_\delta L_{dq} & 2L_\delta^2\\-8L_{dq}^2 & 4L_\delta L_{dq}\end{bmatrix} +R_s\begin{bmatrix}0&2\\0&0\end{bmatrix}\Bigr), \\
M_2 &= -L_{dq}L_{rd}^4\begin{bmatrix}1&0\\0&0\end{bmatrix}, \quad M_3 = 4R_s R_r L_{rd}^2\begin{bmatrix}0&L_\delta\\0&2L_{dq}\end{bmatrix}, \\
M_4 &= 4R_r R_s L_{rd}^2\begin{bmatrix}2L_{dq} & \frac{L_\delta}{2}\\ \frac{L_\delta}{2} & 4L_{dq}\end{bmatrix}, \quad M_5 = -8R_r R_s^2 L_{rd}^2\begin{bmatrix}0&2\\1&0\end{bmatrix}, \\
M_6 &= 48R_r^2 R_s^2\begin{bmatrix}L_{dq} & -L_\delta\\0 & -L_{dq}\end{bmatrix}.
\end{align*}
Real roots $\{\mu_k\}$ are found via companion matrix eigendecomposition and $i^\star = \arg\min_k\{i(\mu_k)^\top R_\mathrm{eff}\,i(\mu_k)+r_\mathrm{eff}^\top i(\mu_k)\}$.

\section{Polynomial Root-Finding in the Forceful Fast Regime}
\label{app:caseC}

We detail the algebraic root-finding procedure for problem \eqref{eq:forceful fast} which arises in the forceful fast driving regime in which constraints~\eqref{eq:P:C3}--\eqref{eq:P:C4} all bind simultaneously.
Under the parametrization~$i_d = i_{s,r}\cos\theta$, $i_q = i_{s,r}\sin\theta$,
the torque constraint~\eqref{eq:P:C5} is linear in~$i_r$.  $b_1(\theta)\,i_r + c_1(\theta) = 0,$ 
with
\begin{align*}
    b_1(\theta) &= L_{rd}\,i_{s,r}\sin\theta, \\
    c_1(\theta) &= (L_d - L_q)\,i_{s,r}^2\cos\theta\sin\theta - T_p,
\end{align*}
both trigonometric polynomials of degree~2.
For $\sin\theta \neq 0$, the rotor current is uniquely determined:
$i_r(\theta) = -c_1(\theta)/b_1(\theta)$. Applying $t = \tan(\theta/2)$ and clearing the resulting rational denominator yields a \emph{degree-4} univariate polynomial $p(t) = \alpha_0 + \alpha_1 t + \alpha_2 t^2 + \alpha_3 t^3 + \alpha_4 t^4,$ with
\begin{align*}
a_0 &= L_{rd}^2\omega^2(T_p + i_{s,r}\psi_q + L_{dq}i_{s,r}^2)^2,\\
a_1 = a_3 &= 4L_{rd}^2 L_q\omega^2 i_{s,r}^2
             (T_p + i_{s,r}\psi_q + L_{dq}i_{s,r}^2),\\
a_2 &= 2L_{rd}^2\bigl[\omega^2\bigl((T_p-L_{dq}i_{s,r}^2)^2
       + i_{s,r}^2\psi_q^2 \\
       &+ 2(L_q^2-L_{dq}^2)i_{s,r}^4\bigr) 
       - 2i_{s,r}^2 V_{\max}^2\bigr],\\
a_4 &= L_{rd}^2\omega^2(T_p - i_{s,r}\psi_q + L_{dq}i_{s,r}^2)^2.
\end{align*}
Note $a_0,a_4\geq 0$ and $a_1=a_3$; real roots are found via
$4\times 4$ companion matrix eigendecomposition
and $i^\star = \arg\min_k\{i(\mu_k)^\top R_\mathrm{eff}\,i(\mu_k)
+r_\mathrm{eff}^\top i(\mu_k)\}$.

\end{document}